\documentclass[a4paper]{article}

\usepackage[T1]{fontenc}
\usepackage[utf8]{inputenc}
\usepackage{lmodern}
\usepackage{titling}

\usepackage[english]{babel}
\usepackage{csquotes}

\usepackage{graphicx}
\usepackage{float}
\usepackage[font=small,labelfont=bf]{caption}
\usepackage[notes,backend=biber]{biblatex-chicago}
\DeclareFieldFormat[online]{citetitle}{\mkbibquote{#1}}

\usepackage{booktabs}
\usepackage{threeparttable}
\usepackage{siunitx}
\usepackage[most]{tcolorbox}
\usepackage{lipsum}

\usepackage{caption}
\usepackage{graphicx}
\usepackage{float}  % for [H] placement
\tcbset{
  figbox/.style={
    colback=black!3,        % light gray background
    colframe=black!25,      % gray border
    boxrule=0.4pt,          % thin border line
    sharp corners,          % squared corners (or ’rounded corners’ if you prefer)
    left=8pt,right=8pt,top=6pt,bottom=6pt,  % inner padding
    enlarge left by=0mm, enlarge right by=0mm,
  }
}

\begin{document}
\title{Google’s hidden empire\\[1ex]
\large New data reveals the company’s unexpected reach and the failure of antitrust authorities wielding Chicago-School economics to address vertical power}
\author{Aline Blankertz,$^{1\ast}$ Brianna Rock, Nicholas Shaxson$^{2}$\\
%\author{Aline Blankertz, Brianna Rock, Nicholas Shaxson}
\normalsize{$^{1}$Rebalance Now, Berlin.}
\normalsize{$^{2}$ Carr-Ryan Center, Harvard Kennedy School.}\\ 
\\
\normalsize{$^\ast$To whom correspondence should be addressed;}\\\normalsize{E-mail: aline.blankertz@rebalance-now.de.}
}

\maketitle

\begin{abstract} 
\noindent
This paper presents striking new data about the scale of Google’s involvement in the global digital and corporate landscape, head and shoulders above the other big tech firms. While public attention and some antitrust scrutiny has focused on these firms’ mergers and acquisitions (M\&A) activities, Google has also been amassing an empire of more than 6,000 companies which it has acquired, supported or invested in, across the digital economy and beyond. The power of Google over the digital markets infrastructure and dynamics is likely greater than previously documented. We also trace the antitrust failures that have led to this state of affairs. In particular, we explore the role of neoclassical economics practiced both inside the regulatory authorities and by consultants on the outside. Their unduly narrow approach has obscured harms from vertical and conglomerate concentrations of market power and erected ever higher hurdles for enforcement action, as we demonstrate using the examples of the failure to intervene in the Google/DoubleClick and Google/Fitbit mergers. Our lessons from the past failures can inform the current approach towards one of the biggest-ever big tech M\&A deals: Google’s \$32 billion acquisition of the Israeli cloud cybersecurity firm Wiz.

\end{abstract}

\section*{Introduction}

In March 2025 Google\footnote{The ultimate parent of Google is Alphabet Inc. The main operating business of Alphabet Inc. is Google, the Alphabet Inc. holding level was introduced in 2015 but the operations remain principally with Google. For the purpose of this paper we refer to Google and the holding level Alphabet Inc. interchangeably.} announced its intent to buy the Israeli-American cloud cybersecurity firm Wiz, in a \$32 billion deal. The companies are now seeking regulatory approval.\autocite{reuters2025google} 
Most large businesses use multiple clouds offered by Amazon, Google, Microsoft and others; Wiz is a widely used platform that enables companies to monitor and protect their “attack surfaces” across these clouds. It has grown spectacularly fast since its founding in 2020: from zero to a \$32 billion valuation in five years, it is a classic tech unicorn.

This deal carries significant risks for the economy and society, however. Google has been found guilty by US courts of “willfully acquiring and maintaining monopoly power” in key markets.\autocite{usvgoogle2025decision}
Cybersecurity needs to be invasive to some degree to monitor real-time activity across all aspects of a businesses, internal and customer-facing. Given Google’s history of pioneering surveillance capitalism, this deal potentially raises significant privacy and other concerns.\autocite{zuboff2023age} %FIXME HM Check is this the source for Zuboff you mean? in the google doc it is just ZUBOFF FN

Public criticism of the deal has been fairly muted so far. This report provides new data and analysis highlighting several often under-appreciated dangers and risks associated with the acquisition.

In Section 1 we explore the regulatory response to past Google acquisitions through two case studies: first, Google’s acquisition of the advertising technology (adtech) company DoubleClick in 2007; and second, the wearable tech company Fitbit, which it acquired in 2021. We focus on the review by the European antitrust authority, that reviewed both deals. In each case the European Commission essentially dismissed extensive concerns raised by a variety of third parties, from both competition and privacy perspectives. Subsequent events have shown that key warnings were prescient and that the Commission’s judgment and analysis dismissing these concerns was mistaken, leading to extensive monopolization of digital markets by Google, and many ensuing harms. In adtech, for example, Google has since comprehensively monopolized the relevant markets, broken promises, broken the law, compromised regulators, and taken surveillance capitalism to new levels.

In Section 2 we deepen this analysis of the bigger Google Empire. After noting Google’s already sizeable (and well known) history of mergers and acquisitions (M\&As), we describe a much larger galaxy of almost 6,000 firms that Google has invested in or has supported in kind, directly or via its own venture capital firms, far more than its big tech peers.\footnote{In this study, “big tech” refers to Alphabet/Google, Amazon, Meta/Facebook, Apple and Microsoft.} 

In Section 3 we analyze the Wiz acquisition in light of the historical regulatory and enforcement failures that we document. We also situate it in the wider societal and geopolitical situation in Europe and other parts of the world that are dependent on US technological infrastructure, as the Trump administration shifts the United States’ stance away from (frequent) co-operation and collaboration towards confrontation and dominance.\footnote{As another example, the US administration established of a “National Energy Dominance Council” in February, 2025.} This includes wielding the market power of US firms like Google in ways that are both economically extractive and geopolitically influential, even coercive.

The conclusion points out avenues for further research and for antitrust authorities to adjust their analytical approaches in order to better capture the business realities of vertical integration and to address the competitive harms from it.

\section{Antitrust authorities’ repeated mistake: \\ 
ignoring the power from vertical integration} 

Before delving into the two case studies, it is instructive to consider the general merger review processes involving Google. The two cases we will discuss are not outliers, but representative of a bigger picture of under-enforcement. Since 2004, only four acquisitions by Google have been reviewed by the European Commission. Three were unconditionally approved (DoubleClick, Motorola Mobility, Photomath) and one (Fitbit) was subject to commitments. Very important transactions were not even reviewed by the European Commission: for example, Google’s acquisition of the British company DeepMind in 2014 that went on to develop Google’s generative AI system Gemini. In total, the number of deals reviewed by the European Commission is less than two percent of the more than 250 acquisitions by Google, over the recorded period. 

One reason for this low figure of deals reviewed is the way that the thresholds that trigger the review are set. For example, thresholds focus on turnover, which might underestimate the importance of fast growing companies. The European Commission can, in collaboration with national jurisdictions, review mergers that do not meet formal thresholds. Yet, even if a review takes place, the use of Chicago School/IO economics results in false negatives, as illustrated below. So, without changing the assessment tools, there seems little benefit in changing the thresholds. 

\subsection{Google’s DoubleClick acquisition: an original sin}

In 2023, the US Department of Justice (DOJ) launched an antitrust investigation into the abuse of Google’s dominance in the online advertising market. Website publishers based in the US sell more than 5 trillion digital display advertisements per year, generating over \$20 billion in annual revenue.\autocite{usdoj2023google_compaint}

According to the DOJ, a Google official in 2016 compared the company’s pervasive power over the advertising technology (“adtech”) stack – the high-tech high-speed and complex digital systems that connect advertisers to online publishers to allow relevant ad placements – to a situation where “Goldman or Citibank owned the NYSE.”\autocite{usdoj2023google_compaint} The possible harms are clear: for example that publishers and website creators earn less, advertisers pay more and quality, including privacy, suffers. Google’s own documents have shown that it keeps at least 35 cents of each advertising dollar that flows through its adtech tools.\autocite[p.~19]{usdoj2023google_compaint} 

In the EU, an investigation was launched in 2021 (and is still ongoing) into Google’s adtech practices, alleging that it is abusing its dominance by favoring its own online display advertising technology services.\autocite{ec2021googleadtech}

The irony of both investigations is that they are repeating past errors, which are by now well known. 

Some 15 years earlier, the US Federal Trade Commission (FTC, the sister authority to the DOJ in antitrust matters) and the European Commission approved without imposing any conditions, the transaction that above all enabled Google to monopolize the adtech stack: its acquisition of the advertising technology company DoubleClick, which already operated on both sides of the online advertising market, serving both publishers and advertisers, and it was setting up an ad exchange between them.\footnote{The DoubleClick transaction was approved in 2007 by the FTC and in 2008 by the European Commission. DoubleClick as the EC decision put it at the time, “mainly sells ad serving, management and reporting technology worldwide to website publishers, advertisers and advertising agencies, in addition to ancillary services. It is launching an intermediation (ad exchange) platform and owns Performics, a search engine management (“SEM”) agency.”}

Before the acquisition, many informed market participants had explicitly warned the EU and US authorities of potentially dramatic consequences post-transaction. Many of these warnings proved prescient,\footnote{See \autocite{epic2007googledoubleclick} which lists an extensive range of concerns, including privacy concerns, filed and published by EPIC, the Center for Digital Democracy, U.S. PIRG and others.} and subsequent events showed that the authorities had got their analyses very wrong.

Instead of listening to sound advice from businesses and others, they approved the DoubleClick transaction using reasoning informed by a form of economic analysis often referred to as Industrial Organization, which by the time of the transaction had become the main toolbox for assessing firm behavior and competition on both sides of the Atlantic.

\begin{tcolorbox}[breakable, colback=black!3, colframe=black, boxrule=0.5pt, title=What is IO Economics?]
Industrial Organization (IO) economics is a branch of neoclassical economics dedicated to studying interactions between companies and consumers. IO economics was heavily influenced by the Chicago School. IO studies firms, markets and their interactions in sanitized and abstract forms, using stylized models and a broad range of often untested assumptions about the shapes of curves on graphs (e.g. supply, demand, cost curves). IO is used by academics and competition authorities, but it is not used by companies in the normal course of business or by corporate analysts to understand and steer business operations or to take business decisions. In this paper we will use the terms Chicago School antitrust economics and IO economics interchangeably, as we consider the way in which IO is currently practiced reflects Chicago School precepts.
\end{tcolorbox}

The DoubleClick acquisition took place after the European competition regime had undergone a series of dramatic changes, the result of sustained US pressure for Europe to fall into line behind the US more corporate-friendly and Chicago-School influenced approach. This pressure had intensified after the European Commission in 2001 blocked the GE/Honeywell merger, a merger of two US companies, that the US authorities had approved. Proponents of the Chicago School then called out European authorities for being unsophisticated, in particular for lacking IO economics expertise, in contrast to the US DOJ which “has a much larger professional staff and employs over 50 PhD economists.”\autocite[p.~35]{grantneven2005ge_honeywell} The US pressure eventually led to the European competition regime undergoing what European Commissioner Mario Monti flagged as a “Big Bang” of reforms in 2004, which brought in a “more economic approach” into EU competition rules and merger control, and the introduction of the post of Chief Economist, to be filled by officials trained in IO economics.\autocite{monti2003eucompetition}

By the time of the Google/DoubleClick transaction, the rise in prominence for IO economics had dramatically weakened EU merger control and enforcement.\footnote{On the softening of merger control, see \autocite{rock2024merger}.} IO-reliant analysis was unsuited to identify the strategies that Google was planning to (and subsequently would) engage, as the following sections explain.

\subsubsection{How IO neutralized vertical concerns}

A central consequence of the “more economic approach” has been the ascendancy of an idea that only “horizontal” mergers between competitors are a concern for competition. It has effectively neutralized concerns about vertical\footnote{Vertical mergers and competition concerns are similar to other non-horizontal concerns stemming from e.g. the accumulation of market power in ecosystems. For ease of reading, we use vertical and non-horizontal interchangeably.} mergers, meaning those involving acquisition of a supplier or customer or an actor in another part of a supply chain or ecosystem. For example, a pesticide manufacturer that acquires another pesticide manufacturer could raise horizontal concerns, while a pesticide manufacturer that acquires a soy and maize seed producer would raise vertical concerns. Google buying DoubleClick, which lay in a different part of the adtech supply chain from its existing assets, was a vertical merger.

The examples below show how the rise to dominance of IO economics on both sides of the Atlantic has enabled the dismissal of concerns raised by competitive rivals to the merging firms based on sound business reasonings. Abstract IO models have been instrumentalized to construct almost insurmountable “IO obstacle courses” for authorities seeking to address vertical concerns.

Consider the work of established professors of IO economics such as Jay Pil Choi discussing the logics of the FTC’s unconditional approval of Google/Doubleclick, in an analysis by FTC-affiliated economists discussing the logics of the FTC’s unconditional approval of Google/Doubleclick:
\begin{quote}
“Although it is possible, as a matter of theory, for a firm possessing substantial market power in one market (“market A”) to extend this market power profitably into another market (“market B”), the conditions necessary for this are stringent (and are necessary, but not sufficient conditions).”\footnote{\autocite[p.~211]{baye2008googledoubleclick}. While Choi is not a co-author of Baye et al., his work is regarded as foundational in building these arguments.}
\end{quote}

They then list five conditions they consider necessary to be able to show that leveraging (vertical) market power should be considered a concern, each a difficult hurdle for enforcers to jump over. The conditions they raise are:
\begin{quote}
(1) The firm must have substantial market power in the market for A; 

(2) The seller of A must credibly commit to bundling A and B together; e.g., by technologically integrating them such that it is very costly to unbundle them; it is not sufficient for the monopolist merely to offer A and B as a bundle if the bundle can be easily unbundled by the firm at a later date; 

(3) The production of B must be characterized by substantial scale economies (so that in equilibrium, only a few firms will be active, leading potentially to imperfectly competitive pricing; and the prospective loss of sales by independent sellers of B from the bundling of A and B will cause a substantial increase in their unit costs, leading to exit/entry deterrence); 

(4) A and B should not be strongly complementary; otherwise the standard “one monopoly profit” critique applies (i.e., the monopolist is better off with competition in the complementary market); and 

(5) The producers of B cannot easily enter the market for A; if they can, then competition will take place in the sale of the bundle, and the attempt to deter entry/induce exit from market B will fail.\autocite{baye2008googledoubleclick}

\end{quote}

Overall, listing the 5 cumulative conditions, the FTC’s analysis dismissed the concern that Google would bundle\footnote{Bundling refers to a strategy where a company commercializes the combination of multiple products at more favorable conditions than when selling the products individually. Google combined DoubleClick services with their ad exchange Google AdX, so the individual services of DoubleClick were not available anymore standalone. Additionally, Google practiced self-preferencing by technically combining its demand side platforms (DSPs) with its ad-stack including DoubleClick.} its online ad business with DoubleClick’s software and discriminate against competitors. But that is what happened. Google did combine DoubleClick’s services with its ad exchange and integrated the services into its unified advertising technology systems. Google engaged in a practice of self-preferencing of its own advertising services as a unified suite, compared to other competitors that relied on Google’s adtech infrastructure elements. 

This illustrates how IO economics generates tremendous problems for public authorities. Making the problem worse, companies under investigation typically hire consultants specializing in IO economics language, which have been accused of “spamming the regulator” with multiple large submissions to overwhelm their capacity to act to uphold the law.\autocite{jugl2023spamming}

In the DoubleClick case, the authorities did not disclose the names of the consultants. However, one economic consultancy, RBB, volunteered that they advised Google in both the FTC’s and European Commission’s reviews.\autocite{rbbecon2008googledoubleclick} While RBB’s actual submission is not available, they summarized their (winning) arguments by introducing a highly theoretical analogy typical for IO arguments that abstract away from relevant features of the markets under consideration. They compare the merger in fast-moving digital markets to a “stylized setting” involving mergers in commodities – the markets for zinc, copper and brass. %We detail their argument in Annex 1 to illustrate how removed these arguments were from the reality of the business of DoubleClick and Google. 
We do not know to what extent RBB’s submissions were accepted; but while the EC’s decision does not refer to the analogy, the conclusions align. 

The European Commission concluded that, first, Google and DoubleClick are not in the same business, so the transaction does not present horizontal concerns. Google sold online ad space while DoubleClick did not; DoubleClick was a leading supplier of ad-serving technology (that ensures ads are placed in the correct places) while Google only offered it as an ancillary service.\autocite{ec2008googledoubleclick} 

Having dismissed the horizontal concerns, the Commission then went on to state how unlikely it would be that post-acquisition Google could leverage its market power through its vertical links with DoubleClick, running its analysis through the gauntlet of the five conditions outlined by the FTC economists to ensure that vertical concerns were dismissed.\footnote{This decision dismissing vertical concerns is very much in the spirit of work by Robert Bork, who led the main charge to neutralize ‘vertical’ enforcement. See \autocite{bork1954verticalintegration}.}

The European Commission made all the following statements—which from today’s perspective are all demonstrably false:\autocite{ec2008googledoubleclick}

\begin{quote}
\begin{itemize}
    \item In the overall online advertising market Google “does not have a sufficient degree of market power to be able to foreclose competitors.”\footcite[para.~334]{ec2008googledoubleclick}
    \item In both search and non-search ad intermediation, Google “most likely lacks the incentive to [foreclose,] both on the advertiser side and on the publisher side.” \footcite[para.~339]{ec2008googledoubleclick}
    \item On the publisher side, the threat of competition would make it “difficult, if not impossible for the merged entity to foreclose its rivals” but it may have limited ability to foreclose by bundling ad space with ad serving.\footcite[para.~345]{ec2008googledoubleclick}
\end{itemize}
\end{quote}

The European Commission concluded that a merged Google/DoubleClick would be unlikely to bundle Google’s search ad offering with DoubleClick’s ad serving technology, because “the large losses that the described bundling strategy would likely produce in the merged entity’s core business, coupled with the limited gains from revenues in ad serving would render any such strategy unprofitable.”\footcite[para.~354]{ec2008googledoubleclick}
 Furthermore, “even in the very unlikely event that the merged entity nevertheless engages in a foreclosure\footnote{Foreclosure means to exclude or disadvantage suppliers or customers.} strategy involving the bundling of Google’s search ad services or (search) ad intermediation services with DoubleClick’s ad serving (and possibly including also non-search intermediation in the bundle), such a strategy would be very unlikely to have a significant detrimental effect on competition.” \footcite[para.~356]{ec2008googledoubleclick}

The European Commission also dismissed privacy concerns relating to the transaction. Google gained access to state-of-the-art advertising technology that significantly enhanced its ability to collect, integrate, and analyze large volumes of personal data for ad targeting purposes. The European Commission dodged the issue, referring in its decision to separate legislation aimed at protecting privacy that would make any concerns about privacy redundant.\autocite[para.~368]{ec2008googledoubleclick}

The FTC’s and Commission’s statements that the merger posed no competitive threat and that they expected vigorous competition to continue, were simply not borne out by events. The authorities’ own subsequent documents confirm how wrong their analysis was.

DoubleClick is deeply integrated in Google’s adtech systems and Google did demonstrably engage in discrimination against other competitors in the access to its adtech, a practice labeled self-preferencing in later antitrust cases.

\subsubsection{Competitive harm resulting from Google/DoubleClick}

In September 2025 the European Commission fined Google €2.95 billion: 

\begin{quote}
   “for breaching EU antitrust rules by distorting competition in the advertising technology industry (‘adtech’). It did so by favouring its own online display advertising technology services to the detriment of competing providers of advertising technology services, advertisers and online publishers.”\autocite{EU_Commission_2025-09-04}  
\end{quote}

 One could argue that times were very different in 2007, when Google/DoubleClick was approved. Neelie Kroes, the Competition Commissioner in charge declared in 2007, a few months before greenlighting the takeover, that “[t]he merger tsunami is a good sign […]. These processes […] must be allowed to run their course without undue political interference.”\footnote{\autocite{Kroes2007CompetitionPolicy}. In retrospect, the statement is worrying on many levels, not least because it encouraged M\&A in an overheated economy ahead of the 2008 financial crisis.}

In 2020, the New York Times looked back at the acquisition of DoubleClick by Google. At that point, it was undeniable that the transaction had been a “game changer” that allowed or at the least accelerated Google’s monopolization of the digital advertising industry. “A growing number of antitrust experts say it’s the sort of deal that should no longer be possible,” the newspaper reported.\autocite{Lohr2020GoogleDoubleClick}

Yet that very year, another important transaction by Google was waved through by regulators on both sides of the Atlantic following the very same faulty Chicago School / IO playbook, dismissing serious concerns that had been raised about the deal.

\subsection{Fitbit: vertical problem, wrong tools}

In December 2020, the European Commission conditionally approved Google’s \$2.1 billion acquisition of Fitbit. On the face of it, the investigation into the transaction could be understood to have learned from the Google/DoubleClick failure because it considered both horizontal and vertical concerns. But due to having been blinded by the Chicago-School rulebook and using misguided IO analysis, the Commission got the functioning of the digital economy in the case wrong, again, failing to address vertical concerns effectively. 

In the US, the DOJ did not oppose the transaction. The Australian competition authority, ACCC first expressed concerns and did not consider the merger commitments set by the European Commission as sufficient to address the issues identified. But after the deal was reached with the European Commission, Google closed the acquisition, disregarding the ACCC.\autocite{Smh2021GoogleFitbitACCC}

The Commission accepted behavioral commitments (limiting Google’s conduct to prevent it from abusing power gained as a consequence of the merger) that were supposed to address a range of horizontal and vertical concerns. 

But the Commission’s analysis and remedies did not anticipate the business realities of what subsequently transpired, post-merger. Notably, the Fitbit product has seen a slow, steady decline, losing monthly active users, generating less revenue, and selling significantly fewer devices between 2020 and 2023/2024.\footnote{Monthly active users fell from 40.2 million in 2020 to 38.5 million in 2023; revenue fell from \$1.13 billion in 2020 to \$0.91 billion in 2024; annual device sales fell from 10.8 million to 6.6 million units in 2023. See \cite{BusinessOfApps2024FitbitStats}. See also \cite{Harding2024FitbitObsolescence}.} Whatever Google’s rationale or strategy here, as far as one can tell nearly five years later, the behavioral commitments are also losing relevance as Fitbit dwindles.

We now focus on the reasoning of the European Commission in its decision.\autocite{EC2020GoogleFitbit}

\subsubsection{Concerns and commitments}

“Our investigation aims to ensure that control by Google over data collected through wearable devices as a result of the transaction does not distort competition,” announced European Commissioner Margrethe Vestager as she launched the in-depth investigation into Google/Fitbit.\autocite{EC2020GoogleFitbitPR}

Here the Commission focused on assessing potential anticompetitive effects in the digital markets related to online advertising, health data, operating systems (OSs) and wearable technology. Although the Commission found that Google and Fitbit were not direct competitors in most of the markets highlighted for potential overlap, it investigated horizontal concerns regarding Google gaining access to Fitbit user data. It considered vertical effects by assessing concerns about Google potentially foreclosing downstream competitors from access to data, to its wearable OS, to Google apps, to Google Play, to Google Search or to Fitbit app stores. It also considered conglomerate concerns about Google leveraging its position in Android into wearables. 

The Commission in clearing the transaction found three concerns that it sought to address with behavioral remedies:

(i) Horizontal data concerns: The Commission found that the merger “would give Google control over an important asset, the Fitbit data, that would further strengthen Google’s dominance in the markets for the supply of online search advertising services.”\autocite[para.~427]{EC2020GoogleFitbit}

The commitment intended to address this concern was data siloing for Google Ads. Google committed to not use Fitbit’s health and fitness data for Google Ads; keep this data separate from its other datasets, and implement a data protection system for at least 10 years.

(ii) Vertical data concerns: The Commission found concerns about Google cutting off providers of digital healthcare services from the Web API\footnote{Application Programming Interface, or API, is technology that allows different software and hardware to talk to each other.} and the Fitbit user data.\autocite[para.~531]{EC2020GoogleFitbit} As a commitment, Google committed to maintaining third party access to Fitbit’s Web API for 10 years.

(iii) Conglomerate OS concerns: The Commission found that there was a risk of Google leveraging its position in its smartphone OS Android to wearables.\autocite[para.~817]{EC2020GoogleFitbit} The commitment intended to address this was Android API Interoperability: Google committed to not charging for access and to not selectively degrading the core interoperability APIs in Android by reducing their functionality with third-party wearables devices or companion apps.

On the one hand, the Commission took vertical concerns seriously and did not simply wave the merger through. On the other hand, however, it turned out that the commitments were tackling concerns that, at least in part, seem to have had little to do with the business strategies behind the merger.

 \subsubsection{Business reality meets inadequate behavioral commitments}
We have a much shorter benefit of hindsight for Google/Fitbit than we do for Google/DoubleClick. Even so, Google’s actions soon after closing provide an indication that strategies other than foreclosure or leveraging into Fitbit’s market had likely driven the deal.

Google seems to have let Fitbit deteriorate in ways that the merger commitments were not suited to prevent. 

At the time of the acquisition of Fitbit commentators expressed concerns about mergers referred to as “killer acquisitions”.\footnote{In May 2020, the OECD hosted a roundtable discussion on killer acquisitions and published a policy paper with the same title shortly after, \autocite{OECD2020ConglomerateEffects}.} In March 2020, Commissioner Margrethe Vestager highlighted in a speech the growing concerns surrounding killer acquisitions, describing them as “attempts to neutralize small but promising companies as a possible source of competition.”\footnote{\autocite{Vestager2020GreenDigital}. Transcript no longer available on the European Commission website. Quotation cited in \autocite{ABA2021KillerAcquisitions}.}
 IO economics pundits embraced the concerns in public speeches and publications, as a way of addressing concerns that their prior practices had failed to identify.\footnote{\autocite{Bell2020KillerAcquisitionsDebate}. One challenge of operationalizing the concerns around killer acquisitions is that the definition in the discussions by IO economists and competition practitioners varies from buying a company to discontinue its product, to buying a competitor to remove a competitive constraint.} 

Against this background, it is surprising that the European Commission, did not identify the “neutralization” of Fitbit as a possible risk of the acquisition by Google. We cannot exclude that this risk was assessed and considered unlikely, but non-public information on the investigation points in this direction. In this case the European Consumer association BEUC did call on the European Commission to investigate whether the Google/Fitbit transaction amounted to a killer acquisition.\footnote{\autocite{BEUC2020GoogleFitbitConcerns}. The report states, “The Commission should address whether the proposed transaction amounts to a “killer acquisition’.”}

Developments after the transaction suggest such concerns were not unfounded. In May of 2021, less than 6 months after the approval of the Google/Fitbit transaction, Google and Samsung announced a strategic partnership where Google would license its wearable OS to Samsung, to compete with Apple in wearable devices.\autocite{KoreaHerald2024GoogleFitbit} Google also launched its own wearable, Pixel Watch, in October 2022, to be maintained alongside Fitbit.\autocite{CNET2022FitbitPixelWatch} It is likely that Google was already developing its own competing product at the time of completing the deal. 

By February 2023, technology news website The Verge ran a subtitle: “Longtime Fitbit users are beyond fed up with server outages, nerfed products, and Google sunsetting their favorite social feature.”\autocite{Verge2023FitbitOutage} In January 2024, reports emerged that the Fitbit co-founders and team members were leaving, and that Google was integrating Fitbit technology with its own smart watch Google Pixel Watch, launched two years after the acquisition of Fitbit, indicating that the value of their contribution to Google was not retained in the medium term.\autocite{Gizmochina2024FitbitExit}

Then, in July 2024, Google discontinued the Fitbit dashboard.\autocite{Heise2024FitbitDashboard} Some users deplored the move: as one commenter put it, “the loss of Fitbit’s large-screen browser-based interface is a shame for the subset of users who rely on it.”\autocite{AndroidPolice2024FitbitDashboard} 

More recently, Google’s flagship generative-AI system Gemini has been deployed on Samsung watches, but there has been no such announcement on Fitbit watches at the time of writing.\footnote{\autocite{Samsung2024WearablesGemini} Our last check on Gemini for Fitbit was on Oct 23, 2025.} Commentators also reported that “Google quietly confirmed Google Assistant’s days on Fitbit watches are numbered” in April 2025.\autocite{AndroidCentral2024FitbitAssistant}

Another indication that Google has not pursued a longer-term strategy on using medical data is that Google announced in October 2025 that it is selling Verily Life Sciences, its subsidiary focused on tools to collect and organize health data to predict disease.\autocite{Bloomberg2025VerilySpinoff} 

All these developments indicate that Google does not have and may not have had serious intentions of generating a relevant business related to health data and wearables. It is hard to speculate why Google purchased Fitbit if it was not planning to develop it. Considerations that may have played a role are the license income from the use of its OS by Samsung, or a threat from Apple had it acquired Fitbit. 

The commitments were intended to prevent Google from pursuing strategies that would harm competition following the merger, but they were not designed in anticipation of what really happened. In addition, the Commission’s decision to accept behavioral commitments was surprising, given the EU revised merger remedies notice expresses a preference for structural remedies: 
\begin{quote}
    “Accordingly, commitments which are structural in nature, such as the commitment to sell a business unit, are, as a rule, preferable from the point of view of the Merger Regulation’s objective, inasmuch as such commitments prevent, durably, the competition concerns which would be raised by the merger as notified, and do not, moreover, require medium or long-term monitoring measures.”\autocite[para.~15]{EC2008RemediesNotice}
\end{quote}

Combining three at least 10-year commitments in a complex and fast-changing digital value chain reflected the Commission’s overconfidence in its understanding of the market dynamics; and in light of this, its approach fell flat. No commitments were in place to ensure Fitbit would remain a viable competitor or to prevent its users from a worsening experience. Instead, behavioral commitments are becoming increasingly irrelevant as Fitbit loses users, making it a less important source of data for digital healthcare providers.

Another factor that casts doubt on the effectiveness of behavioral commitments is that Google and other big tech firms have a bad track record in terms of compliance with EU laws, as the next section demonstrates.

 \subsection{No signs of change away from missing the business reality of vertical integration}

What conclusions might we draw from the DoubleClick and Fitbit cases? In the two large acquisitions by Google assessed by the competition authorities, the authorities relied on convoluted economic reasoning and failed to correctly anticipate the business strategies of Google. The authorities also dismissed complainants using the economic analysis and arguments provided by economic consultancies. One conclusion could be that the antitrust authorities should listen more to complainants, including from civil society and less to economists and economic consultants, given their track record detailed above.

We consider it self-evident that the authorities should reject Chicago School economics reasoning when assessing transactions involving Google, as it systematically underestimates risks related to vertical integration and leveraging of market power across markets. IO economics underestimates risks related to vertical integration and leveraging of market power across markets, and has adapted poorly to digital markets. The introduction of the killer acquisition concept by IO economists may have served to explain why their existing models led to under-enforcement, but it does not seem to have led to better outcomes in practice. 

Vertical integration compounds other challenges in regulating big tech firms, including challenges to compliance and deep pockets. Google and other big tech firms have a bad track record in terms of compliance with EU laws. One illustration is the year-long back-and-forth over Google Shopping and Google Android. Another illustration is the non-compliance findings and further investigations related to the Digital Markets Act and the GDPR.\footnote{For example, \autocite{EC2024DMANonCompliance} See also, \autocite{Ryan2025GuardianRearm}.} The Commission has shied away from considering the sheer size and financial power of companies as a risk to competition when assessing acquisitions by Google. Although the European Commission has at least once in a recent merger case considered the large size of companies as a possible harm to competition, in the 2012 EMI/Universal decision.\footnote{\autocite{EC2012UniversalEMIInvestigation} In this case, the Commission highlighted that the proposed transaction would lead to high market shares.} %Large companies in the digital space will tend to be vertically integrated. 

These concerns related to vertical integration are very likely to get even more pressing as the advent of Artificial Intelligence (AI) boosts the potential for vertical integration through size and deep pockets. 

To illustrate how vertical concerns pervade the AI landscape, investor Keith Rabois “breaks down the AI market through the lens of vertical integration” on the All-In podcast in July 2025: 
\begin{quote}
    “Products really require vertical integration (…) You build a product taking advantage of where in the stack you have the most competitive advantage. But then you leverage that and it reinforces. It’s why Apple, despite missing the AI wave, is still a pretty good company from any empirical standpoint (…) their performance is absolutely miserable in the most important technology breakthrough in the last 70 years. But the company is still alive, and still worth trillions of dollars because it’s vertically integrated.”\autocite{AllIn2025Post}
\end{quote}

So far, the European Commission has not seemed to depart from the narrow theories of harm that systematically failed to stop Google acquisitions. Theories about killer acquisitions do not seem suitable to address under-enforcement. They merely offer a new label for an old problem: that large companies dispose of their rivals through acquisitions that harm competition, leaving the problem unaddressed.

New attempts at legislation show a desire to design new tools, but thus far they remain rooted within an IO framework and result in inadequate provisions. In 2022, the European Union introduced new legislation called the Digital Markets Act (DMA) to moderate the power of the largest tech companies, designated as gatekeepers, a group of companies that includes Google. The legislation includes, for example, obligations of interoperability and prohibits certain self-preferencing practices.\footnote{Interoperability requirements under the DMA are supposed to ensure that hardware and software can communicate effectively so that users can switch between platforms and services. For example, it requires gatekeepers to allow third party apps and hardware to interconnect with theirs.}

While the DMA could have the potential to address some competition concerns, it is not mainly targeted at mergers. It only contains one provision requiring gatekeepers to report all mergers irrespective of whether they meet merger review thresholds. This obligation seems inadequate because the Commission is only informed about, but not automatically tasked with reviewing, these mergers. The inadequacy becomes even clearer when we see that Google has shifted its approach from an acquisition strategy to a more below-the-radar investment strategy, as the next section shows.

\section{Google built a visible and invisible empire through acquisitions and partnerships that escape regulatory scrutiny}

This section moves beyond acquisitions and reviews investments by Google over the past 15 years. The antitrust authorities’ assessments of Google’s acquisitions under represent the company’s significance in the markets where it operates in at least two respects: 

\begin{enumerate}
    \item The number of deals reviewed by the European Commission is less than two percent of the more than 250 acquisitions by Google, as shown in Section 1.
    \item Google’s strategy to invest in companies through its venture capital (VC) arms extends its influence over thousands of market participants, as we outline in this section. 
\end{enumerate}

We find that in the last four years alone Google invested in over 2{,}500 companies, many more than the other GAFAM companies combined. 
It created a network of investees (mostly in software for businesses and other customers) over which Google can have influence, which is unaccounted for by antitrust authorities.

Competition regulators are solely investigating acquisitions, if at all, while Google has shifted towards a strategy of investing without acquiring full control. On all metrics, Google has increasingly resorted to investments over the past years, more than other big tech companies. 

We conducted a review of the acquisitions and investments by Google, as well as by Amazon, Apple, Meta (previously Facebook) and Microsoft, to allow for a comparison among the so-called GAFAM companies. The data was collected from PitchBook, a data provider for private capital markets, including VC and startup deals. Our data captures more than the deals reported by the data providers specializing in public markets, such as Eikon or Bloomberg. Our review includes transactions reported by companies that PitchBook identified as subsidiaries of each parent company, thus in principle including the main subsidiaries of Alphabet (Google). For completeness, we list the scope of the legal entities in our review of GAFAM from 2014-2024 at the end of the next section.

\subsection{Google’s shift in investment strategy from acquisitions to minority investments and support in kind}

Financial data providers, including PitchBook, classify transactions as mergers typically if the companies report it as such, corresponding to a change of control. Investments, by contrast, are acquisitions of minority stakes in companies, for example, 15 percent of voting shares. PitchBook also assigns an investment classification for grants and provision of support in kind, such as through accelerators that help startups find other investors.\footnote{In PitchBook’s methodology, “Accelerator/Incubator refers to an event in which a company joins a program that variably provides funding, office space, technological development and/or mentorship, often in exchange for equity in the company.” See \autocite{PitchbookDealTypes}.} The participation in accelerator programs provides companies that run the programs both access to information on the innovation, and an opportunity to shape the business plans of startups.

Figure 1 shows that while Google’s number of acquisitions decreased over the past 15 years, the number of investments over the same period rocketed. Acquisitions and investments were at roughly similar levels in 2010 (34 vs. 20), but then Google’s investments took off, rising more than sixfold in the ten years to 2022, reaching a peak of 743 investments that year. They have since remained stable at over 600 additional investments per year.

\begin{figure} [H]
  \begin{tcolorbox}[figbox]
\caption{Deals announced by Google in the form of acquisitions and investments, 2010--2024}
\renewcommand{\textfraction}{0.5}
\includegraphics[width=1\textwidth]{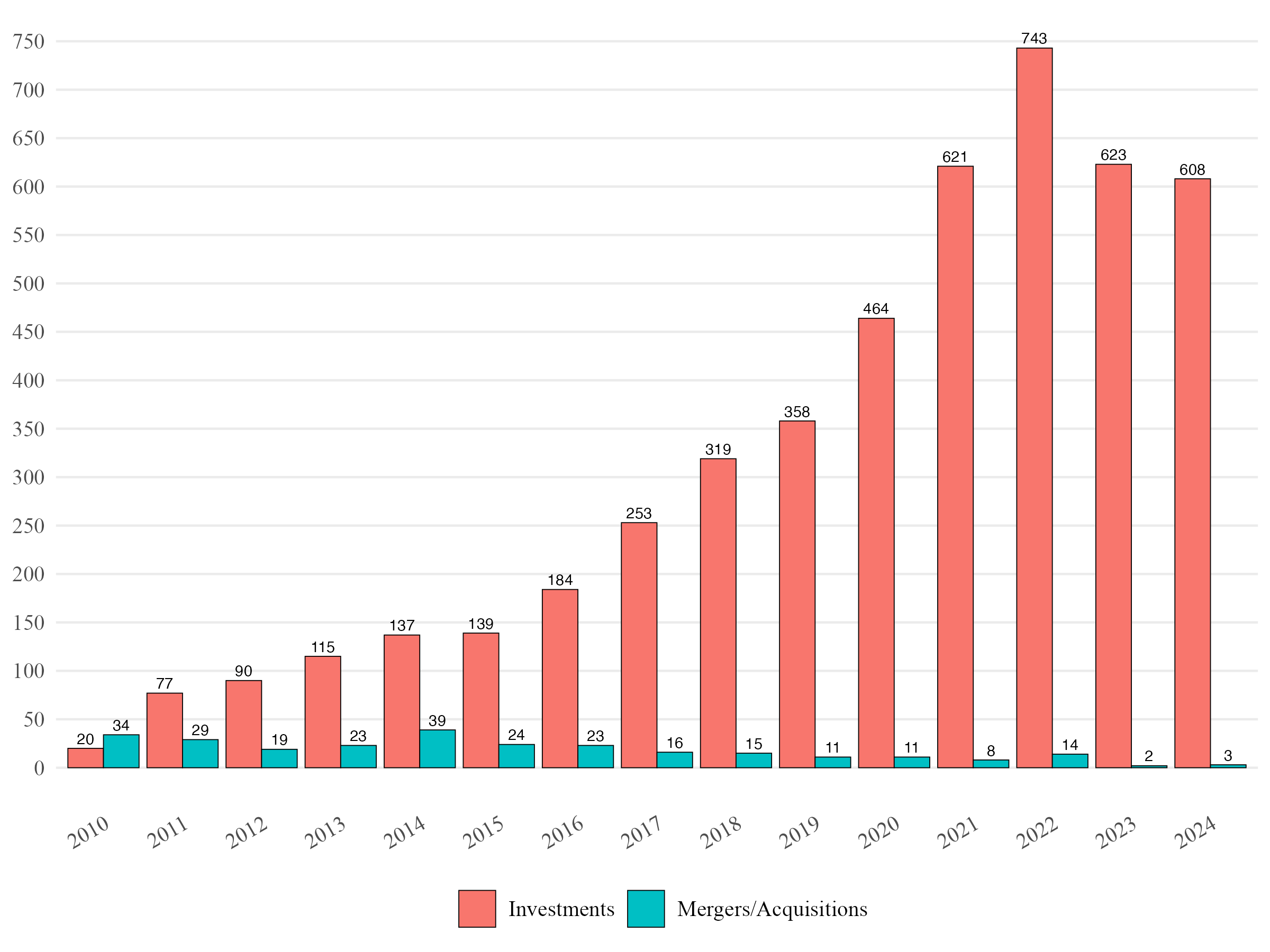}
\footnotesize {\\[2pt]Source: PitchBook data extracted 3Q 2025. 

Scope of Data: All companies listed as affiliated by PitchBook that have reported an investment in another company as of the extraction date. Deals with undisclosed dates were excluded.}
 \end{tcolorbox}
\end{figure}

Figure 2 shows that the increase in investments by Google corresponds primarily to the investments of its VC arm Google for Startups. Between 2015 and 2024, Google for Startups alone accounted for 57.8\% of all Google’s investments. 

\begin{figure} [H]
  \begin{tcolorbox}[figbox]
\caption{Google’s investments by affiliate firm, 2010--2024}
\renewcommand{\textfraction}{0.5}
\includegraphics[width=1\textwidth]{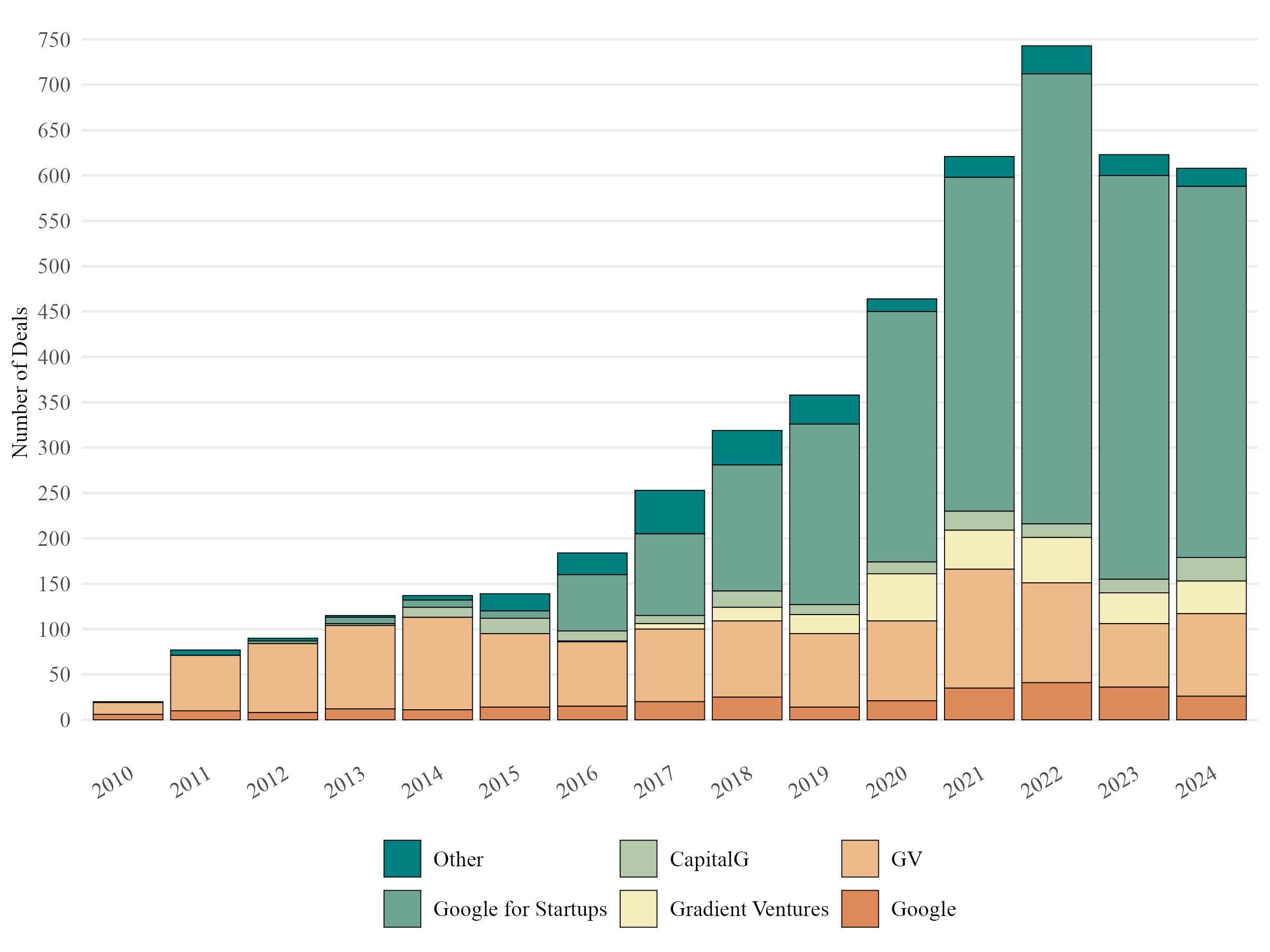}
\footnotesize {\\[2pt]Source: PitchBook data extracted 3Q 2025. 

Scope of Data: All companies listed as affiliated by PitchBook that have reported an investment in another company as of the extraction date. Deals with undisclosed dates were excluded.}
 \end{tcolorbox}
\end{figure}

This active investment into startups policy is not discussed in competition cases, although it could greatly impact the direction of innovation by those startups and, therefore, by the markets where Google is active overall. 

Google for Startups provides support to startups in different forms, including grants, that do not give Google access to company’s voting shares but can result in loyalty and partnerships, such as data sharing. Many firms that join Google for Startups programs do not receive equity, but rather support in kind. For example, Google has a program whereby the company grants to startups free credits to use its cloud. The beneficiaries of this program might partly overlap with the investments we capture. However, no data are available on how many companies received free Google cloud credits without any other form of support. This suggests that the numbers presented above are a lower bound of the number of start-ups that are directly dependent on Google’s support.

The scale of Google’s investment into other businesses in the form of capital or in kind does not come through in many existing studies. For example, in 2021 the FTC published a study of acquisitions by big tech firms over a 10-year period from 2010 to 2019.\autocite{FTC2021TechPlatformStudy} The five respondents reported 819 total transactions, on top of transactions that had been reported to the FTC or the DOJ for merger review during the same period.\footnote{The study focused on transactions that have not been reported for review to the antitrust authorities under the Hart-Scott-Rodino (HSR) Antitrust Improvements Act.} The figure corresponds to an average of approximately 164 transactions per company, presented in the FTC study, though data on individual companies was not presented. Additionally, the study specified that approximately 50 additional smaller transactions below \$1 million were excluded as well as 160 financial investments, and patent acquisitions below \$2.5 million. This threshold would explain why the study omitted many Google investments. 

The fact that big tech companies resort to investments and support in kind for startups has been reported by researchers.\autocite{Rikap2024AIGovernanceBeyondOwnership} For example, Cecilia Rikap focused on Amazon, Google and Microsoft for their strategies of “control beyond ownership” (in relation to AI) and points to the use of cloud credits specifically. In the data reported by Rikap, Google appears as the most active investor in AI startups among the three companies she analyzed. However, our data below show an even more pronounced distance between Google and Microsoft/Amazon.\footnote{See Table 1 below. See also \autocite{Rikap2024AIGovernanceBeyondOwnership} 2024; and \autocite{Rikap2024IntellectualMonopolies}}

\subsection{Google out-invests in the acquisition of influence in other businesses}

We now compare the activities of Google to that of other big tech companies: in Figure 2 (above) for acquisitions, and in Figure 3 (below) for investments. The two graphs confirm Google’s shift to investments.

Figure 3 shows that the acquisitions of all GAFAM companies decreased from 2014 to 2024. In 2014, they each acquired between 9 and 39 companies; by 2024, this fell to between 0 and 5. This fall is the most pronounced for Google (from 39 to 3), while Meta’s merger activities remained fairly stable between 4 and 9 companies per year, falling to only 2 in 2023 and 0 in 2024.

\begin{figure} [H]
  \begin{tcolorbox}[figbox]
\caption{Merger \& Acquisition deals initiated by GAFAM companies, 2014--2024}
\renewcommand{\textfraction}{0.5}
\includegraphics[width=1\textwidth]{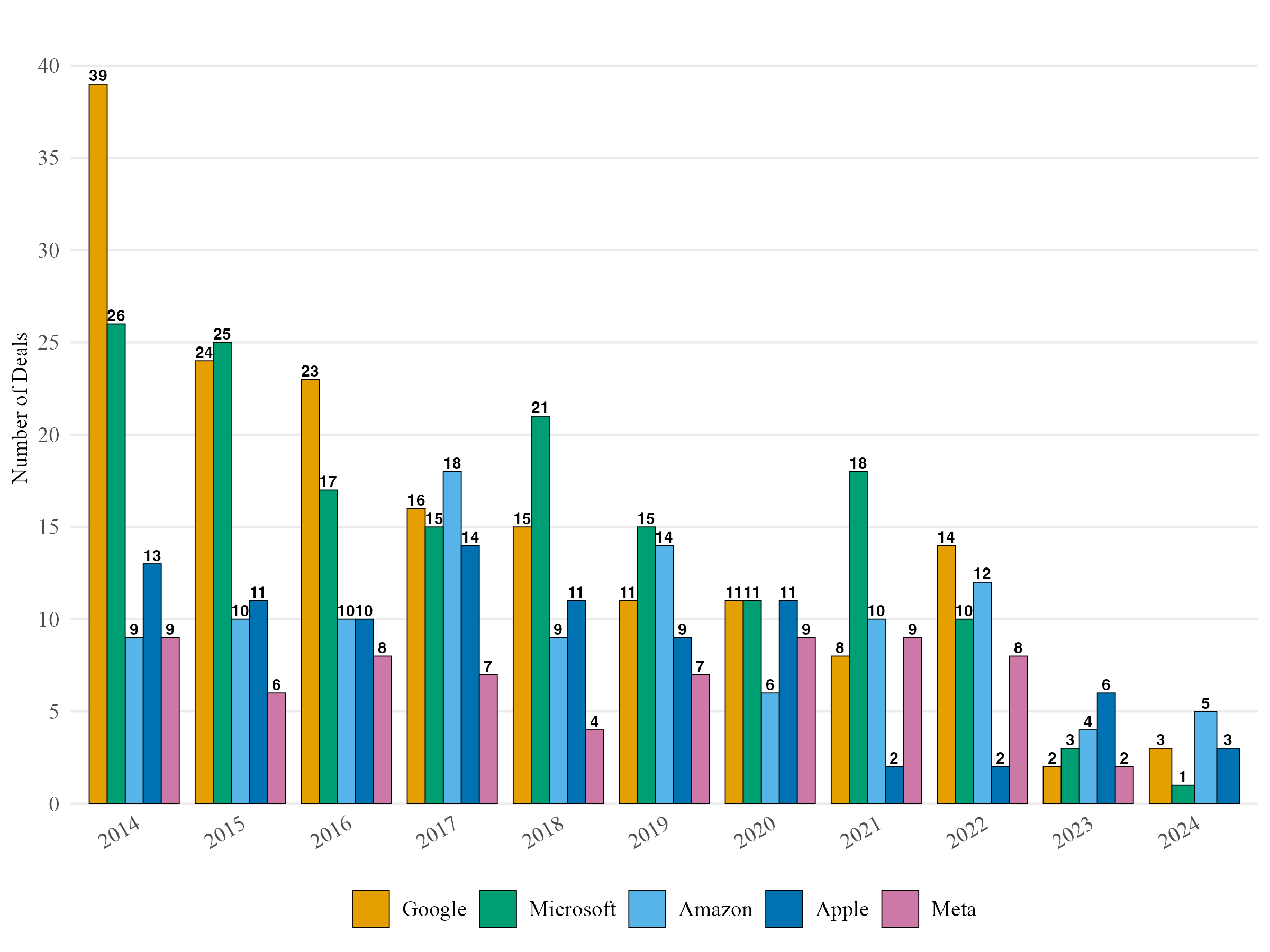}
\footnotesize {\\[2pt]Source: PitchBook data extracted 3Q 2025. 

Scope of Data: All companies listed as affiliated by PitchBook that have reported an investment in another company as of the extraction date. Deals with undisclosed dates were excluded.}
 \end{tcolorbox}
\end{figure}

This change might be explained by a number of reasons. It is plausible that the relative ease of pursuing other strategies such as investments, might have played a role in increasing their control over networks of technology. Companies might have factored in increasing attention from antitrust authorities, which could have led to more interventions had mergers continued at the same level. At the same time, the growth of the digital sector overall is likely to have generated more opportunities for investment, with many start-ups seeking funding.

Figure 4 shows that the shift to investments was pursued only by some GAFAM companies in the past 10 years, with Google from 2017 emerging as the clear leader. Apple and Meta invested in fewer than 50 companies per year throughout the period. Amazon increased its investments by a factor of 19, albeit from a low level of 12 per year in 2014 to 227 in 2024. Both Google and Microsoft started from a relatively high level in 2014, with Microsoft investing in more companies per year than Google from 2014-2016, averaging 186 a year, after which its investments fall to 90-164 per year. Google, by contrast, pursues a huge expansion of its investments; in each year from 2018 it has invested into more companies than the other GAFAM companies combined.

\begin{figure} [H]
  \begin{tcolorbox}[figbox]
\caption{Investment deals announced by GAFAM companies, 2014--2024}
\renewcommand{\textfraction}{0.5}
\includegraphics[width=1\textwidth]{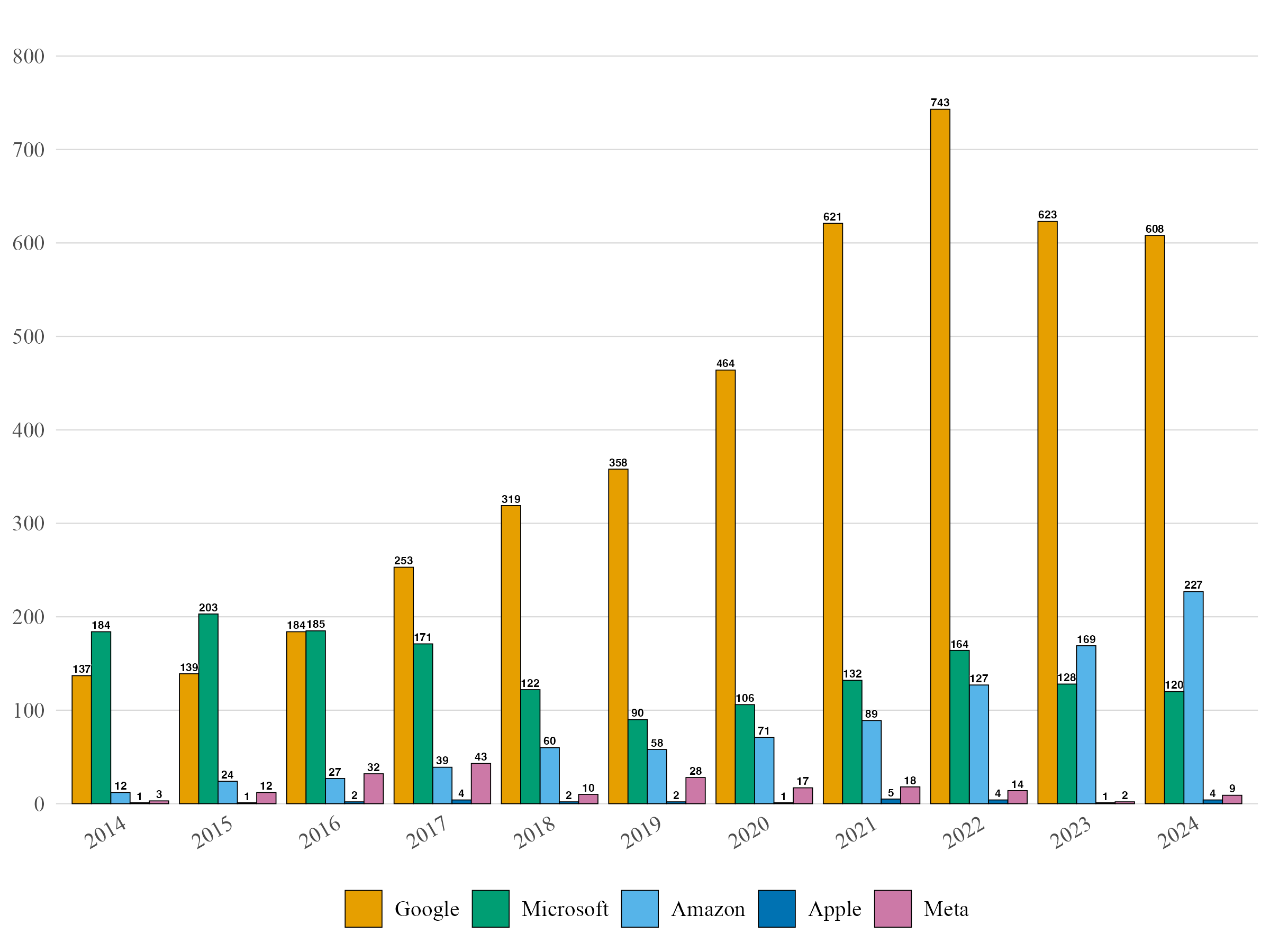}
\footnotesize {\\[2pt]Source: PitchBook data extracted 3Q 2025. 

Scope of Data: All companies listed as affiliated subsidiaries by PitchBook that have reported an investment in another company as of the extraction date. Deals with undisclosed dates were excluded.}
 \end{tcolorbox}
\end{figure}

It is beyond the scope of this article to assess the reasons for these different strategies. There are also other strategies for extending control beyond ownership, which are not reflected here: for example, investing in open-source ecosystems that allow GAFAM companies to shape industry standards; another example is the cloud credit agreements mentioned above.\autocite{Rikap2024AIGovernanceBeyondOwnership} This information calls for further research to breakdown the investments by type. This information is not readily available for many of the transactions we have reviewed.

\begin{tcolorbox}[breakable, colback=black!3, colframe=black, boxrule=0.5pt, title= Note: List of subsidiaries included in the analysis for each company]

For each GAFAM company, we have included all companies listed as their affiliated subsidiaries by PitchBook. Reported deals that did not contain date information were excluded from our analysis.
\\
In Figures 3 and 4 our scope of analysis includes the GAFAM and all companies listed as their affiliated subsidiaries by PitchBook that have reported an investment in another company as of the extraction date. The legal entities included in these analysis for each GAFAM are listed below.
\\
\textbf{Amazon:} Alexa Accelerator, Amazon, Amazon Alexa Fund, Amazon Catalyst, Amazon future Engineer, Amazon Industrial Innovation Fund, Amazon Smbhav Venture Fund, Amazon Web Services, ART19, Audible, AWS Startups, Axio (Financial Services), Black Business Accelerator, eero, LoveFilm International, MGM Studios, One Medical (Clinics/Outpatient Services), Ring, Souq, The Climate Pledge, Twitch Interactive, Veeqo, Whole Foods Market, Wondery, Zoox. 

\textbf{Apple:} Apple, Beats Electronics, Shazam. 

\textbf{Google:} Adopt A Startup, BrightBytes, CapitalG, Firebase, Google, Google Assistant Investment, Google Cloud Platform, Google DeepMind, Google Digital News Initiative Innovation, Google Fiber, Google for Startups, Google Foundation, Google.org, Gradient Ventures, GV, Mandiant, Sidewalk Labs, Verily Life Sciences, Youtube. 

\textbf{Meta:} Facebook Community Accelerator, FbStart, Instagram, Meta Platforms, New Product Experimentation, WhatsApp.

\textbf{Microsoft:} Activision Blizzard, Encore Business Solutions, Github, Grupo-imagine, LinkedIn, M12, Metaswitch Networks, Microsoft, Microsoft 4Afrika, Microsoft AI Innovate, Microsoft Climate Fund, Microsoft Cloud \& Mobile Application Incubator, Microsoft for Startups, Microsoft Payments (Malaysia), Microsoft’s Your Big Idea, Minecraft, Nuance Communications, Rare (Entertainment Software), Simplygon, Skype, The Kinect Accelerator, Yammer, ZeniMax Media.

Access to PitchBook and Bloomberg is possible through Harvard Library. 
\end{tcolorbox}

Investment deals represented in Figures 1, 2 and 4 are those for which the announcement date is known, 8,385 in total for the big tech firms. PitchBook also reports transactions where no date is known, 1,171 investments, creating a more rounded picture. Table 1 reports that over the past 15 years, Google announced 5,899 investments into different companies. Second was Microsoft with 2,246, followed by Amazon with 1,096. Meta invested in 267 companies over the same period and Apple in just 48, for a total of 9,556, as Table 1 shows. 

\begin{table}[ht]
  \begin{tcolorbox}[figbox]
  \renewcommand{\arraystretch}{1.2} % row height
  \setlength{\tabcolsep}{10pt}      % column padding
  \centering
  \begin{threeparttable}
    % FLUSH-LEFT CAPTION 
    \captionsetup{justification=raggedright,singlelinecheck=false}
    \caption{Total GAFAM investment records through 2024}
    \label{tab:gafam-dated-vs-nodate}

    % RIGHT-ALIGNED NUMERIC COLUMNS 
    \begin{tabular*}{\linewidth}{@{\extracolsep{\fill}} l
      S[table-format=4,table-alignment=right]
      S[table-format=4,table-alignment=right]
      S[table-format=4,table-alignment=right]}
      \toprule
      & \multicolumn{1}{c}{\textbf{No date}} 
      & \multicolumn{1}{c}{\textbf{Dated}} 
      & \multicolumn{1}{c}{\textbf{Total}} \\
      \midrule
      Amazon     & 78   & 1018 & 1096 \\
      Apple      & 2    & 46   & 48   \\
      Google     & 781  & 5118 & 5899 \\
      Meta       & 29   & 238  & 267  \\
      Microsoft  & 281  & 1965 & 2246 \\
      \addlinespace
      \textbf{Total} & \textbf{1,171} & \textbf{8,385} & \textbf{9,556} \\
      \bottomrule
    \end{tabular*}
    % SOURCE NOTES 
    \begin{tablenotes}[flushleft]
      \footnotesize
      \item {Source}: PitchBook data extracted 3Q~2025. 
      \item {Note}: “Dated” are investments with a reported date. “No date” are records lacking deal-date information and are therefore excluded from time-series figures.
    \end{tablenotes}
  \end{threeparttable}
   \end{tcolorbox}
\end{table}

Additional sources confirm Google’s oversized presence as a financial investor. Figure 5 below shows that in the first half of 2025, the company GV Management Company, L.L.C., doing business as Google Ventures, is among the top ten largest investors of all deals reported by Bloomberg globally, by number. (Bloomberg does not cover private markets with the same granularity as PitchBook; the investments reported by Bloomberg correspond to larger existing businesses.) 

In Figure 5, Bloomberg shows Google’s GV ventures among 9 other VC firms. No other tech firm makes an appearance in the top investor list, which tends to confirm the greater degree of activity of Google as an investor, compared to other big tech companies. Google’s GV features in 8th place, with 46 deals announced in the first half of 2025. This is just below Khosla Ventures, a VC run by Keith Rabois cited in Section 1 (on the importance of vertical integration). Google for Startups does not appear in the ranking, although, as Figure 2 shows, it provides most of Google’s support for startups. The reason for its absence from the Bloomberg list is that Google for Startups’ equity support is often undisclosed and in many cases Google does not provide direct funding but support in kind, including cloud credits.\footnote{Such deals are too small to be captured by Bloomberg, which focuses on public markets. PitchBook is specialized in private markets and also reports smaller investments.}

\begin{figure} [H]
  \begin{tcolorbox}[figbox]
\caption{First half 2025 all M\&A and investment deals reported by Bloomberg globally, ranked by count}
\renewcommand{\textfraction}{0.5}
\includegraphics[width=1\textwidth]{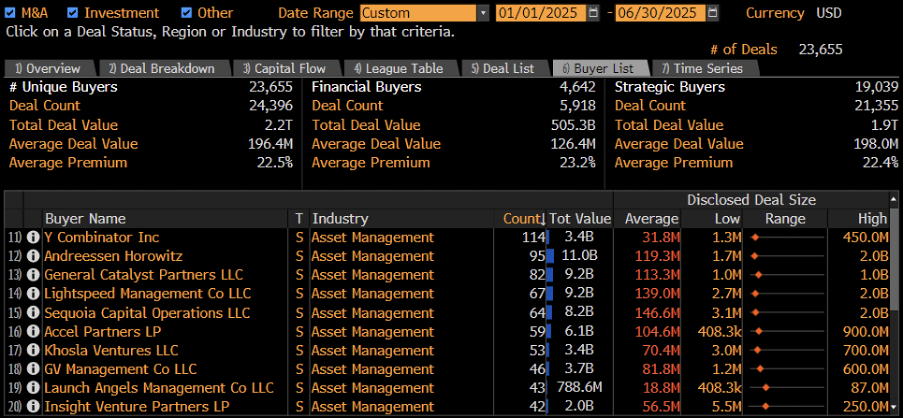}
\footnotesize {\\[2pt]Source: Bloomberg. 

Note: Bloomberg describes GV Management Company, L.L.C. as doing business as Google Ventures and as a venture capital firm.}
 \end{tcolorbox}
\end{figure}

This provides further evidence that Google’s investing activity is out of step with that of other large tech companies, which are seen as Google’s peers. The ubiquitous presence of Google in the economy and innovation space increases Google’s clout also vis-à-vis the US government, as the next section discusses. 

\section{As Google becomes an arm of US geopolitical power, the proposed Wiz acquisition is a test case for vertical enforcement}

This section discusses how the urgency of enforcement is still growing by the day. Google’s influence and power are no longer limited to economic markets, but increasingly extend into the political sphere, threatening to undermine the capacity of governments to take action. Google portrays its cloud and AI technology to the US government as must-have to lead at a geopolitical level. This creates a situation in which the government could feel compelled to use their resources to strengthen Google rather than to control and limit it. Google’s proposed acquisition of Wiz can be seen as a test case for this emerging dynamic.

\subsection{The interests of tech companies merge with those of the US government}

The challenge of regulating Google has never been greater. Due to lobbying and its central place in the digital economy, the interests and power of Google now increasingly resemble \textit{raison d’état} for the United States, and are leveraged to exert power and dominance over others. A statement from the sitting US Commerce Secretary Howard Lutnick in March 2025 illustrates the stance: 

\begin{quote}
    “I say, I want you to give me, you’re going to build for America, you’re going to build the greatest customs processing ever. [...] I say, build it for me for free. I put it in for free. I don’t know what other countries in the world you think are going to buy now. [...] Remember, you have to connect to me. So every country is gonna buy. That’s a great business model. […] Then I go to the heads of Google and Microsoft and Amazon. They’re all for America, building for us.”\autocite{AllIn2025LutnickInterview}
\end{quote}

This is context for our data on Google’s investments, which goes beyond what was already well known about Google’s political clout in Brussels, London, Washington DC, and elsewhere. Extensive accounts exist of its lobbying and influence activities, many of which are hidden.\footnote{See \autocite{TechTransparencyProject2025ShellGame}, which documents not only extensive lobbying activities but also its efforts to cloak those lobbying efforts. See also Lobbyfacts.eu, which recorded nearly €6m a year in officially registered lobbying activities in Europe since 2020.} For example its former CEO, Eric Schmidt has since leaving the company taken up various advisory assignments with the US government. He has repeatedly warned of an AI arms race between the US and China – and many have taken this great-power competition to argue in self-serving ways, notably to push for US government subsidies for big tech firms, and to oppose regulation of AI, which is a critical “dual use” set of technologies serving both commercial and military purposes.\footnote{He has served as Chair of the National Security Commission on Artificial Intelligence established from 2018 to 2021 under the Biden administration, and funded and strongly influenced the White House Office of Science and Technology Policy. See The real scandal behind billionaire Eric Schmidt paying for Biden’s science office. See \autocite{Kim2022SchmidtOSTP}; \autocite{Kim2022SchmidtFrontierFund}; \autocite{Kitchen2023BigTechChina}; \autocite{BrysonMalikova2021AIColdWar}}

These geopolitical factors will likely dampen authorities’ resolve to oppose Google’s proposed acquisition of Wiz. 

Yet the case for action is strong. The proposed Wiz acquisition has some similarities to Google’s DoubleClick acquisition in 2008, and to other acquisitions by Google that were waved through by competition authorities. It poses serious and specific risks of monopolization of the cloud industry by Google, as the next section explains.

\subsection{Google’s Wiz acquisition as a way to increase its market power}

Currently, Google is not the largest player in cloud services. Worldwide business spending on public cloud services\footnote{A public cloud is a third-party cloud like Amazon’s or Google’s, as opposed to a private cloud which is a business’ in-house computing cloud.} was estimated at nearly \$600 billion in 2024, with the Big Three US firms enjoying a nearly two-thirds market share, according to industry estimates: Amazon Web Services (AWS) had 30-32 percent, Microsoft’s Azure had 21-23 percent, and Google Cloud just 11-12 percent.\footnote{See \autocite{Gartner2024CloudSpendingForecast}; \autocite{SRG2025CloudMarket}; \autocite{Hava2024CloudMarketShare}}

However, Google’s smaller market share may underestimate its importance for several reasons. First, it is growing faster than its main peers in cloud, so its market share is growing.\footnote{The global cloud market has grown by 45\% since 2022, from \$228bn to 330bn. \autocite{SRG2025CloudMarket}. While Google’s cloud revenues have grown by 64\% over the same period, from \$26bn to \$43bn. \autocite{Alphabet2025_10K}. Microsoft’s and Amazon’s cloud revenue growth have each averaged 15\% a year from 2021-2024; Google’s has grown by an average annual 22\%. \autocite{Microsoft2025_10K}; \autocite{Amazon2025_10K}} Second, it is more deeply embedded than its peers in the start-up ecosystem and the innovation space, as Section 2 shows. 

Most businesses use a multi-cloud strategy, not least for resilience purposes, but this poses technical challenges for cybersecurity.\autocite{SecurityMag2021MultiCloud} Wiz addresses this by providing “connective tissue” in the cybersecurity realm allowing businesses to have a view across their different clouds and products, in real time on one dashboard. Wiz is seen as intuitive and powerful, and is well respected by businesses and the cybersecurity community.\footnote{See \autocite{TrustRadius2025WizReviews}} 
%\footnote{See \autocite{Reddit2024WizExperience} and \autocite{TrustRadius2025WizReviews}} 

Originally, the founders of Wiz described a stance of ensuring neutrality across different clouds, even calling their offering the “Switzerland of cloud security.”\autocite{InsightPartners2023WizCofounders}

Yet with its Wiz acquisition, Google may be able to shift Wiz away from a “neutrality” stance. Cyber experts talk of “left of boom” and “right of boom” (the “boom” being a cyber event, like an attack.) Left of boom means proactive and preemptive cybersecurity measures to protect “attack surfaces” and prevent attacks, such as through threat intelligence, security training, and so on. “Right of boom” means measures taken after an event, including incident response. Google has recently acquired three other cybersecurity companies that are more “right of boom”: Chronicle, Siemplify, and Mandiant. Of the three transactions, Mandiant was the largest with an acquisition price of \$5.4 billion. The acquisition did not raise any objections by the DOJ and was not formally reviewed by the European Commission.

Wiz would complement them by being more left of boom, thus giving Google a more comprehensive one stop shop offering in the cybersecurity space. 

Investors have described Wiz acquisition as part of a “Trojan horse” strategy to boost its market power over essential online infrastructure.{\footnote{The term “Trojan horse” was used by Chamath Palihapitiya, speaking on the \textit{All-In} tech-business podcast, (\autocite{AllIn2025Episode220}). He added: “When the deal was done, one of the things that Wiz said is, hey, don’t worry, we’re not going to prefer GCP [Google Cloud Platform.] It was interesting because I thought, wait, people actually thought that they would just prefer GCP? No, it’s the exact opposite. Because now what you have are your tentacles. Think of any other service that any cloud provider provides where most of your revenue is coming from the other clouds. Can you name one? Well, it didn’t exist until now.”}} Wiz’s comprehensive multi-cloud offerings may thus help Google expand its market power vertically into new domains. 

Furthermore, Wiz would give Google an intrusive and enduring presence in other businesses’ internal systems. The extensive corporate data it can obtain will be of a very different nature from Google’s mostly user-obtained data e.g. from Google Search: adding huge troves of (new) data is potentially both a competition and surveillance issue.\footnote{\autocite{Bryson2021DataCollapse} Even if Google were forced to disgorge its datasets to competitors to level the playing field, new triangulation via Wiz’s real time data could soon give it a competitive advantage over rivals again.} 

In conclusion, Google’s proposed acquisition of Wiz presents vertical concerns, similar to the situation in the Google/DoubleClick acquisition many years ago. If regulators rely on the same theoretical IO economics as they did before, which is singularly ill-equipped to deal with vertical issues, they risk to fail again to identify the true harms from the proposed acquisition. 

In the US, authorities are currently reviewing the transaction; as of October 2025, we have no information on the EU’s plans to review it. There is considerable geopolitical momentum against intervention in both the US and the EU. At the same time, governments and competition authorities have good reasons for blocking the transaction and refusing to give Google any more power than it already wields, assuming that those governments have customer and citizen interests in mind.

\section*{Conclusion}

We show that Google is building an unexpectedly large empire with geopolitical implications, yet barely challenged by competition authorities. We identify two key reasons for this failure. First, the widespread application of IO economics has blinded competition authorities to effectively analyzing and addressing harms from vertical integration, as was the case in various Google mergers before the European Commission. Second, Google pursues below-the-radar strategies to expand and cement its dominance, including by creating a huge network of dependent start-ups that it provides with funding or other support.

The mistakes made in competition policy have not just allowed Google to amass significant market power in the past: a failure to learn from these mistakes will allow Google to expand its empire further. 

While it is beyond the scope of this paper to develop a comprehensive competition policy agenda, our analysis points in a number of directions that require further research and policy analysis.

The past mistakes indicate a need for different analytical tools that, in addition to their immediate benefits, can help improve legislation. Specifically, there are approaches that can overcome the capture of economic analysis by IO. The tools from accounting and financial analysis (AFA) are better suited to ground the analysis of vertical integration and conglomerate market power in business realities.\footnote{See \autocite{ProMarket2025ExcessivePricing} and \autocite{BEP2024IOFinancialAnalysis}} Using AFA would make it easier to shift enforcement away from currently ineffective approaches such as behavioral remedies and towards structural interventions such as merger prohibitions. 

For example, the Google adtech cases that are pending both in the US and in the EU stem significantly from the failure to intervene in the Google/DoubleClick merger years earlier; requiring Google to divest relevant parts of its adtech business would help address the harm caused by that failure. 

The ongoing expansion of the Google empire that currently escapes review calls for updated legislation and procedures. One option would be for the Commission to conduct systematic reviews of investments and agreements for assessing the market power of Google in any ongoing or future competition cases. 

The dynamic nature and the increasing political weight of big tech companies mean that governments and regulators are likely to come under even more pressure to allow further expansion of market power, with many negative economic, political and geopolitical implications. There is no time to lose in addressing these risks. The proposed Google/Wiz merger would be an important test case to start enforcing properly.

\section*{Acknowledgments}
The authors declare that they have no conflicts of interest or financial relationships to disclose. Nicholas Shaxson is a fellow at the Carr-Ryan Center, he is expressing his personal views, not engaging the Carr-Ryan Center. 
Thanks to Columbia University and Hertie School initiative on Digital Governance for Democratic Renewal for motivating this work. Thanks also to Anna Marchese for comments. The resources used, specifically PitchBook and Bloomberg data can be accessed through Harvard Library.

\clearpage

\clearpage
\printbibliography
\end{document}